\documentstyle[epsfig,amssymb,aps]{revtex}

\begin{document}

    \draft \title{Meanfield treatment of Bragg scattering from a
  Bose-Einstein condensate.}

\author{P.~B.~Blakie and R.~J.~Ballagh} \address{Department of
Physics, University of Otago, P.~O.~Box~56, Dunedin, New Zealand.}
\date{\today}
 
\maketitle
\begin{abstract}
A unified semi-classical treatment of Bragg scattering from Bose-Einstein 
condensates is presented. The formalism is based on Gross-Pitaevskii equations
driven by a classical light field, and leads to a single component equation
incorporating the effects of meanfield nonlinearity and spatial inhomogeneity.
Three dimensional numerical simulations of this equation in cylindrical
symmetry are used to investigate Bragg scattering for a number of cases.
Scattering from a condensate released from a trap produces characteristic
cycling of the atomic momentum, and an analytic description is given using
a linear model incorporating spatial nonuniformity. Simple expressions obtained
for the momentum packet cycling are shown to
accurately describe the full nonlinear behaviour within a well specified
validity range and a transition linewidth is derived. 
For the case of a trapped condensate, a numerical study
of momentum spectroscopy is carried out and the shift of the Bragg resonance
condition due to meanfield nonlinearity is investigated. 

\end{abstract}


\section{Introduction}

The phenomenon of Bragg scattering, where a wave interacts with an extended
periodic structure, was first observed by W. H. and W. L. Bragg \cite{bragg12}
in experiments scattering X-rays from crystal planes.  In 1988 Martin et al.
\cite{martin88} demonstrated the matter wave equivalent using an atomic beam
of neutral atoms interacting with a light grating, and 
subsequently, Bragg  scattering has been widely used
in atomic optics and interferometry  (e.g see \cite{berman97}).

Bose condensates, with their extended spatial coherence, offer significantly
enhanced scope for the application of this technique,  and it has already
become an important tool in the investigation and manipulation of this new
state of matter.  In recent experiments Bragg scattering has been used on 
condensates to engineer the meanfield wavefunction \cite{kozuma99}, to measure coherence 
and correlations \cite{hagley99,stenger99,stamper99},  
and act as a beam-splitter \cite{deng99,torii00}. Most of the existing theory of atomic Bragg
scattering has concentrated on treating plane-wave single atom states (e.g see
\cite{martin88,bernhardt81,oberthaler99,giltner95,adams94}), and while this
captures some key behaviour, it does not account for some distinctive
properties of condensates, such as nonlinear collisional self interaction and
nonuniform  spatial distribution.  Zhang and Walls \cite{zhang94} have
developed a theory of Bragg scattering for cold atom beams which includes some
effects of atomic interactions, although the resulting nonlinear interaction
term they derive is of an unconventional form.  Their accompanying numerical
calculation for a beam-splitter process shows that the inclusion of
nonlinearity can give rise to qualitatively new effects.

In this paper, we present a unified semi-classical treatment of Bragg
scattering from condensates, in which the condensate is described in the
meanfield limit, and the applied laser fields are treated as classical fields.
Beginning with Gross-Pitaevskii equations we obtain a single component
equation describing the interaction of a condensate with a light field
grating formed at the intersection of two far detuned laser beams.  The lasers
may be pulsed or cw, and arbitrary geometries are allowed.  
This formalism is sufficiently general to include diffraction (where the light grating has 
rapid transverse spatial variation or the interaction time is short) and channeling (where the
lasers have high intensity), but in this paper we concentrate on the regime of Bragg
scattering, for which the output state has momentum components only in a single narrow range.
Accordingly we consider only those cases for which the transverse spatial
laser profiles vary slowly in comparison to the condensate, and the interaction time is long.

 In order to characterise the effect of nonlinearity and spatial
nonuniformity, we have numerically simulated our single component 
equation in three dimensions. We consider the case where the light fields are
applied immediately after the condensate is released from a trap, and also the
case where the condensate remains trapped. Some aspects of the behaviour,
notably the shift of the Bragg resonance condition, are due directly to the
nonlinearity. However, we also find that over a usefully wide parameter range,
much of the behaviour can be understood in terms of a framework provided by
the linear results. To facilitate comparison with the linear case, we present
a novel treatment of linear Bragg scattering for a released condensate, which
conveniently incorporates spatial nonuniformity. This model is based on a
partitioned representation of momentum space and under appropriate validity
conditions, reduces to a set of two state systems, allowing simple expressions
to be found for the dynamical behaviour of momentum wave packets, and the
momentum transition linewidth. We determine the validity range over which the
linear treatment accurately describes Bragg scattering of nonlinear released
condensates, and identify two main mechanisms responsible for the failure of
the linear model; meanfield expansion and nonlinear dispersion. We verify our
analysis by numerical simulation of the full nonlinear equations, and also
investigate numerically the nonlinear shift of the Bragg resonance in a
trapped condensate.

The paper is organised as follows. In section II we begin with Gross-Pitaevskii 
equations and derive the one component equation
that is the basis for the rest of the paper. In section III we present an
analytic treatment of linear Bragg scattering from a released condensate,
based on a partitioned representation of the momentum wavefunction, and formulate
the treatment as
a tridiagonal matrix equation. Under well defined approximations, this reduces
to a two-state form which is easily solved, and in section IV we examine the
properties of the solution and introduce dispersion curve plots to help
visualise the implications of our approximations. In section V we consider in
detail the effects of  nonlinearity on Bragg scattering from released
condensates, and finally in section VI, we consider momentum spectroscopy from
condensates in harmonic traps.	

\section{Formulation}
\label{formalism}
We consider as our model a coherent system of atoms in a light field
grating provided by two crossed laser beams.  The laser fields are
treated as plane-wave classical fields with wave-vectors and
frequencies $ {\bf k}_{i} $ and $ \omega _{i} $ respectively. 
For convenience we shall assume the electric fields are of equal
intensity, and so the total field is given by
\begin{equation}
{\bf E}_{T}({\bf r},t)=\frac{1}{2}{\bf E}_{0}(t)\Lambda ({\bf r},t),
\end{equation}
 where
\begin{equation}
\Lambda ({\bf r},t)=[e^{i({\bf k}_{1}\cdot {\bf r}-\omega
_{1}t)}+e^{i({\bf k}_{2}\cdot {\bf r}-\omega _{2}t)}+c.c].
\end{equation}
We have neglected any spatial variation of the slowly varying amplitude $ {\bf E}_{0}(t) $ 
 (although it would be trivial to include), but include a time dependence to allow for 
 the possibility of pulsed fields. The condensate is treated in the
 meanfield limit and has two internal states $ |g\rangle $ and $
 |e\rangle $ separated by a Bohr frequency $ \omega _{eg} $ and
 with a dipole matrix element $ {\bf d} $.  The coupling of theses
 two states by an electric field is characterised by the Rabi
 frequency, $ \Omega _{0}(t)={\bf d}\cdot {\bf E}_{0}(t)/\hbar $. 
 The meanfield equations for the ground and excited state
 wavefunctions $ \psi _{g} $ and $ \psi _{e} $ are the time
 dependent Gross-Pitaevskii equations

\begin{eqnarray}
i\hbar \frac{\partial \psi _{g}}{\partial t} & = & -\frac{\hbar
^{2}}{2m}\nabla ^{2}\psi _{g}+V_{Tg}({\bf r},t)\psi_g-\frac{1}{2}\hbar \Omega _{0}(t)\Lambda
({\bf r},t)\psi _{e}\label{full_evolve_eq} \\
 & + & w_{gg}|\psi _{g}|^{2}\psi _{g}+w_{eg}|\psi _{e}|^{2}\psi
 _{g},\nonumber \\
 & & \nonumber \\
i\hbar \frac{\partial \psi _{e}}{\partial t} & = & -\frac{\hbar
^{2}}{2m}\nabla ^{2}\psi _{e}+\hbar \omega _{eg}\psi
_{e}+V_{Te}({\bf r},t)\psi_e-\frac{1}{2}\hbar \Omega ^{*}_{0}(t)\Lambda ({\bf r},t)\psi
_{g}\label{full_evolve_EQ2} \\
 & + & w_{eg}|\psi _{g}|^{2}\psi _{e}+w_{ee}|\psi _{e}|^{2}\psi
 _{eg},\nonumber
\end{eqnarray}
where $ w_{gg} $ and $ w_{ee} $ characterise the intra-species
collisional interactions, and $ w_{eg} $ the interspecies
interaction. 
The functions $V_{Ti}$ represent the trapping potentials for the respective internal states,
and we include time dependence in order that they can be dynamically altered. 
 We assume the laser frequencies $ \omega _{i} $ are sufficiently off resonant from the
$ |g\rangle \rightarrow |e\rangle $ transition that the laser field 
undergoes negligible modification in propagating through the condensate, and
furthermore that spontaneous emission (and its effect on condensate coherence) can be
ignored. In this large detuning limit it is also permissible to adiabatically eliminate
the excited state from Eq. (\ref{full_evolve_eq}),
leaving only an equation for $ \psi _{g} $.  Details of this
procedure are given in the Appendix \ref{adiaelimsec}, where it is shown that provided
the detuning $ \Delta $ ($ =\omega _{1}-\omega _{eg} $) is sufficiently large that the energy
$\hbar\Delta$ exceeds all other external energies associated with the excited state, we
obtain the one component equation for the ground state wavefunction
\begin{equation}
\label{spatial-evolution-equation}
i\hbar \frac{\partial \psi }{\partial t}=-\frac{\hbar ^{2}}{2m}\nabla
^{2}\psi+V_T({\bf r},t)\psi+V(t)\lambda ({\bf r},t)\psi +w|\psi |^{2}\psi . 
\end{equation}
The wavefunction $ {\psi}=\exp(-i\int_0^tV(s)ds)\,\psi_g,$ is the ground state
wavefunction in an interaction picture, and for clarity we have written  $ w $
for $ w_{gg} $ and $V_T({\bf r},t)$ for $V_{Tg}({\bf r},t)$, and have defined the quantities

\begin{eqnarray}
V(t) & = & \frac{|\Omega _{0}(t)|^{2}}{2\Delta },\\
\lambda ({\bf r},t) & = & \cos (\Delta {\bf k}\cdot {\bf r}-\delta
\, t),
\end{eqnarray}
where $ \Delta {\bf k}={\bf k}_{1}-{\bf k}_{2} $ and $ \delta
=\omega _{1}-\omega _{2} $.  
The function $ \lambda ({\bf r},t) $
contains the periodic spatial and temporal evolution of the optical potential, and arises 
by neglecting certain rapidly rotating terms in the adiabatic elimination (see Appendix \ref{adiaelimsec} for
details).

 For the coupled system described by Eq. (\ref{spatial-evolution-equation}) the duration of the grating amplitude 
$V(t)$ is a key factor in determining the nature of 
the system response. For Bragg scattering the pulse duration $T_p$ must be 
of order \cite{keller99}
\begin{equation}
T_p > {m\over \hbar|\Delta {\bf k}|^2}.
\end{equation}

In most of this paper (sections \ref{analytictreatmentsec}-\ref{appnonlinear}) we will consider 
the case of  Bragg scattering from a condensate released from a harmonic trapping potential 
(i.e. the trapping potential is set to $0$ at $t=0$). It is also 
possible to have the trap on during the Bragg pulses, and obtain a
result qualitatively similar to scattering from a released condensate,
provided the trap does not remain on too long. The time duration should be
less than a trap period, (so that the trap itself does not significantly alter 
momentum components) and less than the characteristic time of momentum spreading  of a
released condensate (which occurs due to mean field expansion, see section \ref{appnonlinear}).

\subsection{Numerical Solutions}

\label{dimensionless_unit_sec}

Equation (\ref{spatial-evolution-equation}) is the basic equation for
this paper and is familiar in atom optics.  In general it is difficult
to solve because of the nonlinear term and in order to explore the
types of behaviour that can occur, we have obtained numerical
solutions for a range of parameters.  Our solutions are in three
dimensions, but restricted (by computational resources) to the case of
cylindrical symmetry, a choice which requires $ \Delta {\bf k} $ to
be in the axial direction (i.e. $ \Delta {{\bf k}}=\Delta k\,
\hat{{\bf k}}_{{\bf z}} $).

\begin{figure} 
\begin{center}
\epsfbox{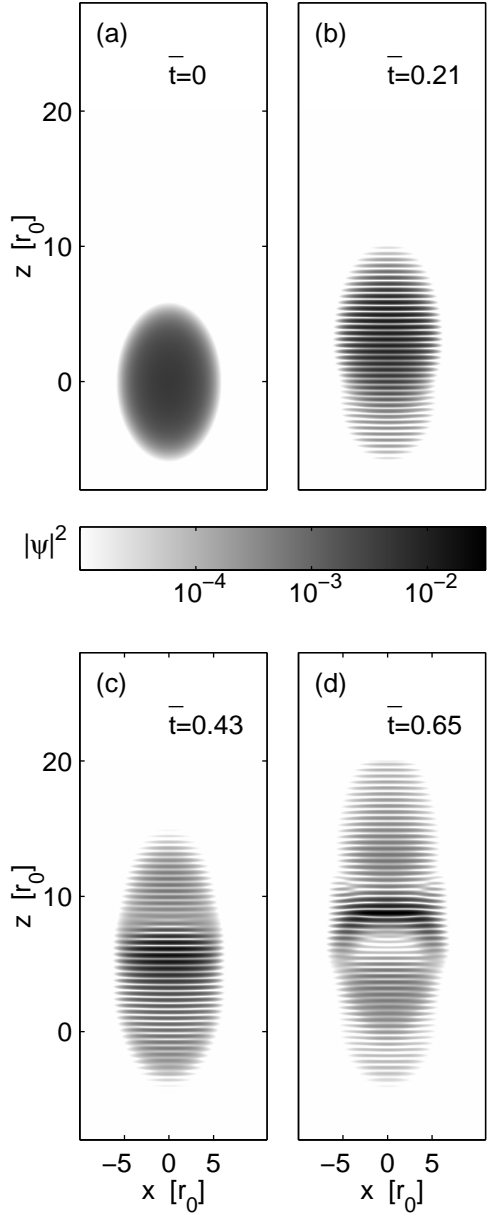}
\end{center}
\caption{\label{complex_density_fig} Probability density (in the
$ x  $-$ z
 $ plane) of a condensate evolving in the presence of a light
grating.  The condensate is initially in the ground state of a
spherical harmonic trap, and at time $ t=0
 $, the trap is switched off and a cw-light grating applied. 
Parameters are $ w=2500\, w_{0}  $,
$ \Delta {\bf k}=14/r_{0}  $ in the
$ z  $-direction, $
\delta =196\, \omega _{T}  $ and $
V=20\, \omega _{T}  $.}
\end{figure}

We present in Fig.  \ref{complex_density_fig} a sequence of images
showing the density of a condensate evolving in the presence of a 
light grating, and illustrating the complexity that can typically occur. 
The condensate was released from a spherical harmonic trap at time $t=0$, and the 
light grating applied immediately after. During this simulation we can
see fringes developing in the density profile, which is characteristic
of interference between a stationary and a moving wavepacket.  As time
progresses the spatial distribution extends in the $ z $ direction
at a much faster rate than spreading occurs in the perpendicular
directions, and the density distribution becomes less uniform (see
Fig.  \ref{complex_density_fig}(d)) with distinct regions of high and
low density apparent.

%figure1

For notational simplicity our numerical solutions are given in
dimensionless units.  Since the initial states for the simulations we
discuss in this paper are related to eigenstates of harmonic traps, we
have chosen to use harmonic oscillator units for the
problem.  The transformation between S.I ($ \{t,{\bf r},{\bf
k},\delta ,w\} $) and harmonic oscillator units ($
\{\bar{t},\bar{{\bf r}},\bar{{\bf k}},\bar{\delta },\bar{w}\} $) is

\begin{eqnarray}
t=\bar{t}/\omega _{T}, &\quad & {\bf r}=\bar{{\bf r}}\, r_{0},\\
{\bf k}=\bar{{\bf k}}/r_{0}, &\quad & \delta =\bar{\delta }\, \omega
_{T},\\
w=\bar{w}\, w_{0}, &
\end{eqnarray}
 where $ \omega _{T} $ is the trap frequency, $ r_{0}=\sqrt{\hbar
 /2m\omega _{T}} $ and $ w_{0}=\hbar\omega_Tr_0^3$.

\section{Analytic Treatment of the Linear Case}
\label{analytictreatmentsec}

It is evident that the simplified treatment of Bragg scattering of
condensates encompassed in Eq.  (\ref{spatial-evolution-equation}) can
give rise to quite complex behaviour.  Here we derive an analytic
treatment of the model that will provide a simple understanding of the
behaviour over a wide parameter regime.  We begin by transforming the
equation into momentum space, where the momentum space wavefunction $
\phi ({\bf k},t) $ is defined as the spatial Fourier transform

\begin{equation}
\phi ({\bf k},t)=\frac{1}{(2\pi )^{\frac{3}{2}}}\int \psi ({\bf
r},t)\, e^{-i{\bf k}\cdot {\bf r}}\, d^{3}{\bf r}.
\end{equation}
Henceforth, for the sake of notational clarity, we will suppress the
explicit time dependence of the momentum wavefunction, and simply
write \textbf{$ \phi ({\bf k}) $}, for \textbf{$ \phi ({\bf k},t)
$}.  Straightforward manipulation of Eq. 
(\ref{spatial-evolution-equation}) leads to

\begin{eqnarray}
i\hbar \frac{\partial \phi ({\bf k})}{\partial t} & = & \frac{\hbar
^{2}|{\bf k}|^{2}}{2m}\phi ({\bf k})+ \hbar V(t)[\frac{1}{2}\phi ({\bf k}+\Delta {\bf k})e^{i\delta
t}+\frac{1}{2}\phi ({\bf k}-\Delta {\bf k})e^{-i\delta
t}]\label{full-momentum-eq} \\
 & + & \frac{w}{(2\pi )^{3}}\int \int \phi ({\bf k}_{1})\phi
 ^{*}(-{\bf k}_{2})\phi ({\bf k}-{\bf k}_{1}-{\bf k}_{2})\, d^{3}{\bf
 k}_{1}\, d^{3}{\bf k}_{2},\nonumber
\end{eqnarray}
and we can see that the effect of the light grating is to linearly
couple states of momentum $ {\bf k} $ to neighbouring states
\textbf{$ {\bf k}\pm \Delta {\bf k} $}.  It is useful to partition
momentum space to incorporate this regularity, in an analogous way
that Brillouin zones are used in solid state physics to reflect the
periodicity of the crystal lattice.  Choosing the $ \Delta {\bf k} $
in \textbf{$ {\bf k} $}-space to be along the $ {\bf z} $
direction, we divide the $ \hat{{\bf k}}_{{\bf z}} $ axis into
intervals of length $ |\Delta {\bf k}| $, and label them by the
integer $ n $, such that the centre of the $ n^{\hbox {th}} $
interval (along the $ \hat{{\bf k}}_{{\bf z}} $ axis) is at
\textbf{$ k_{z}=n\Delta k $}, (and $ n=0,\pm 1,\pm 2,\pm 3,\ldots
$).  The wavefunction $ \phi ({\bf k}) $ can now be re-expressed as
a set of wavefunctions $ \tilde{\phi }_{n}({\bf k}) $, each defined
only within the $ n^{\hbox {th}} $ interval of the partitioned $
{\bf k} $-space, i.e.
\begin{equation}
\tilde{\phi }_{n}({\bf k})\equiv \phi ({\bf k})\qquad \{{\bf k}:\,
(n-\frac{1}{2})|\Delta {\bf k}|<k_{z}\leq (n+\frac{1}{2})|\Delta {\bf
k}|\}.
\end{equation}
 Within each interval, the momentum vector is uniquely defined by its
 offset from the central $ {\bf k} $ value, and thus we finally
 replace the wavefunction $ \phi ({\bf k}) $ by the set of
 partitioned wavefunctions
\begin{equation}
\label{phases_on_binned_wfns_eqn}
\phi _{n}({\bf q})=\tilde{\phi }_{n}({\bf k})\, e^{-in\delta t},
\end{equation}
 where ${\bf q}={\bf k}-n\Delta {\bf k},$
 i.e. the domain of $ \phi _{n} $ is all values of $ {\bf q} $
 which have a $ \hat{{\bf k}}_{{\bf z}} $ component in the range $
 [-|\Delta {\bf k}|/2,|\Delta {\bf k}|/2] $.  We have included phases in
 the definition in Eq.  (\ref{phases_on_binned_wfns_eqn}), for later
 convenience.  The utility of these new wavefunctions arises because
 $ \phi _{n}({\bf q}) $ and $ \phi _{n+1}({\bf q}) $ represent the
 full momentum wavefunction at adjacent positions in $ {\bf k}
 $-space separated by exactly one momentum kick $ \hbar \Delta {\bf
 k} $.  In terms of this new set of wavefunctions, Eq. 
 (\ref{full-momentum-eq}) can be written as the set of equations
\begin{eqnarray}
i\hbar \frac{\partial \phi _{n}({\bf q})}{\partial t} & = & \hbar
\omega _{n}({\bf q})\phi _{n}({\bf q})+\frac{1}{2}\hbar V(t)[\phi _{n-1}({\bf
q})+\phi _{n+1}({\bf q})]\label{binned-evolution-equation} \\
 & + & \frac{w}{(2\pi )^{3}}\int \int \sum _{i,j}\phi _{i}({\bf
 q}_{1})\phi _{-j}^*({\bf q}_{2})\phi _{n-i-j}({\bf q}-{\bf q}_{1}-{\bf
 q}_{2})\, d^{3}{\bf q}_{1}d^{3}{\bf q}_{2}\nonumber
\end{eqnarray}
 where
\begin{equation}
\label{omega-eq}
\omega _{n}({\bf q})=\frac{\hbar |{\bf q}+n\Delta {\bf
k}|^{2}}{2m}-n\delta .
\end{equation}

\subsection{Noninteracting Atoms }

We derive an analytic solution for Eq. 
(\ref{binned-evolution-equation}) by considering the case where
collisional interactions in the BEC are negligible, i.e. $ w=0 $.  We
shall find that this allows us to make a good representation of the
full equation for a large regime of interest for condensates.  Putting
$ w=0 $ allows us to write the evolution equation
(\ref{binned-evolution-equation}) as a linear system:

\begin{eqnarray}
i\frac{\partial }{\partial t}\left[ \begin{array}{c} \vdots \\
\phi _{n-1}({\bf q})\\
\phi _{n}({\bf q})\\
\phi _{n+1}({\bf q})\\
\vdots
\end{array}\right] =\left[ \begin{array}{ccccc}
\ddots & \ddots & 0 & & \\
\ddots & \omega _{n-1}({\bf q}) & \frac{V(t)}{2} & 0 & \\
0 & \frac{V(t)}{2} & \omega _{n}({\bf q}) & \frac{V(t)}{2} & 0\\
 & 0 & \frac{V(t)}{2} & \omega _{n+1}({\bf q}) & \ddots \\
 & & 0 & \ddots & \ddots
\end{array}\right] \left[ \begin{array}{c}
\vdots \\
\phi _{n-1}({\bf q})\\
\phi _{n}({\bf q})\\
\phi _{n+1}({\bf q})\\
\vdots
\end{array}\right] & \label{matrix_evolution_eq}
\end{eqnarray}

\subsection{First Order Couplings and the Two-state
Model\label{Two_state_section}}

The matrix in Eq.  (\ref{matrix_evolution_eq}) displays the couplings
between the partitioned momentum wavefunctions.  For a given momentum
$ {\bf q} $, the effectiveness of the coupling between $ \phi
_{n}({\bf q}) $ and $ \phi _{n+1}({\bf q}) $, is determined by the
size of the coupling $ V $ relative to the momentum detuning,
\begin{equation}
\label{detuning_def_eq}
\Delta _{n}({\bf q})=\omega _{n+1}({\bf q})-\omega _{n}({\bf q}).
\end{equation}
In particular if (for some ${\bf q}_r$), $ \Delta _{n}({\bf q}_r)=0 $, then the corresponding
transition of a particle from momentum state $ |{\bf q}_r+n\Delta {\bf
k}\rangle \rightarrow |{\bf q}_r+(n+1)\Delta {\bf k}\rangle $ is
resonant.  Noting that the momentum kick given to the atoms is $
\hbar \Delta {\bf k} $, (the momentum difference of the photons from
the two light fields) and that
\begin{equation}
\label{Effective_Detuning_Eq}
\Delta _{n}({\bf q})=\frac{\hbar }{2m}\left[ (2n+1)\Delta {\bf
k}+2{\bf q}\right] \cdot \Delta {\bf k}-\delta ,
\end{equation}
then the resonance condition $ \Delta _{n}({\bf q}_r)=0 $ is
recognised as a Bragg condition for the process: that is both momentum
and energy are conserved.

The typical initial condition for Eq.  (\ref{matrix_evolution_eq}) is
a localised momentum wavepacket.  If $ |\Delta {\bf k}| $ is
significantly greater than the width of the wavepacket, the initial
profile is almost completely contained within a single momentum
interval, and thus we can take the initial momentum wavefunction to
have only one of the partitioned wavefunctions $ \phi _{0}({\bf q})
$ to be non-zero.  It is easy to see from Eq. 
(\ref{Effective_Detuning_Eq}) that if one of the $ \Delta _{j}({\bf
q}) $ is very small, then all the others are larger by order $
|\Delta {\bf k}|^{2} $.  For example, if $ \Delta _{0}({\bf q}_r)=0 $
(a first order Bragg resonance), then

\begin{equation}
\label{other-state-detunings-eq}
\Delta _{n}({\bf q}_r)={\hbar \over m}\left[ n|\Delta {\bf
k}|^{2}\right].
\end{equation}
Thus for the case $ \Delta _{0}({\bf q}_r)=0 $, if the neighbouring detunings 
are much larger than $V$ over the entire ${\bf q}$-domain (see also section \ref{2statebreakdown}), 
we can make the secular approximation
\cite{cohen75} that only the coupling $ \phi _{0}\rightarrow \phi
_{1} $ is significant and the evolution equation can be reduced to:

\begin{equation}
\label{two-state-matrix-eq}
i\frac{\partial }{\partial t}\left[ \begin{array}{c} \phi _{0}({\bf
q})\\
\phi _{1}({\bf q})
\end{array}\right] =\left[ \begin{array}{cc}
\omega _{0}({\bf q}) & \frac{V(t)}{2}\\
\frac{V(t)}{2} & \omega _{0}({\bf q})+\Delta _{0}({\bf q})
\end{array}\right] \left[ \begin{array}{c}
\phi _{0}({\bf q})\\
\phi _{1}({\bf q}) \end{array}\right] .
\end{equation}

This equation represents a continuous set of two-state Rabi problems,
where the members of the set are labelled by the variable $ {\bf q}
$.  The solution is easily found in terms of the eigenvectors of the
coefficient matrix in Eq.  (\ref{two-state-matrix-eq}) (e.g see \cite{cohen77}).

\subsubsection{Separable Wavefunction}

The complete Bragg scattering problem in the linear case ($ w=0 $)
is described by Eq.  (\ref{matrix_evolution_eq}), a system of
equations in which the momentum argument $ {\bf q} $ is a three
dimensional vector.  This three dimensional character remains even
when the equations can be reduced to a set of two-state problems, as
in the previous section.  It is possible, however, to find a solution
for the full linear case which simplifies the geometrical character. 
If we assume the partitioned wavefunction is separable, i.e.

\begin{equation}
\phi _{n}({\bf q})=\xi (q_{x})\zeta (q_{y})\Phi _{n}(q_{z}),
\end{equation}

and we take the momentum kick to be in the $ z $ direction, then Eq. 
(\ref{matrix_evolution_eq}) transforms to a set of independent
equations,
\begin{eqnarray}
i\hbar \frac{\partial \xi (q_{x})}{\partial t} & = & \frac{\hbar
^{2}q_{x}^{2}}{2m}\xi (q_{x}),\label{separated-eqn} \\
i\hbar \frac{\partial \zeta (q_{y})}{\partial t} & = & \frac{\hbar
^{2}q_{y}^{2}}{2m}\zeta (q_{y}),\\
i\hbar \frac{\partial \Phi _{n}(q_{z})}{\partial t} & = & \hbar \left[
\frac{\hbar (q_{z}+n\Delta k)^{2}}{2m}-n\delta \right] \Phi
_{n}(q_{z})\\
 &  & +\frac{1}{2}\hbar V(t)\left[ \Phi _{n-1}(q_{z})+\Phi
 _{n+1}(q_{z})\right]. \nonumber
\end{eqnarray}
The wavefunction $ \Phi _{n}(q_{z}) $ obeys an equation identical to
Eq.  (\ref{matrix_evolution_eq}), except that the momentum argument
is now simply the scalar $ q_{z} $.  The wavefunctions $ \xi
(q_{x}) $ and $ \zeta (q_{y}) $ describing the $ q_{x} $ and $
q_{y} $ momentum behaviour are freely evolving one dimensional
wavepackets, and for example with a gaussian initial condition have a
well known analytic solution (e.g \cite{cohen77}).  Reduction of the
system of equations for $ \Phi _{n} $ to a two-state system would
proceed exactly as in the previous section.  We have chosen, in that
derivation, to retain the slightly more general form $ \phi _{n}({\bf
q} $), because it also includes the case of non-separable
wavefunctions, and we will retain the general form $ \phi _{n}({\bf
q} $) in the remaining sections of this paper for the same reason. 
We emphasize however that our results can be transformed to the
somewhat easier separable case by simply making the transformation $
\phi _{n}({\bf q})\rightarrow \Phi _{n}(q_{z}) $.

\section{Results and Features of the Two-State Model}
\label{twostate}
In this section we present the characteristic features of the two
state model and compare its behaviour to the numerical solutions of
Eq.  (\ref{spatial-evolution-equation}) for the case of $ w=0 $.  We
then discuss the parameter regimes for which the two-state model
provides a good description of the full $ w=0 $ behaviour.

\subsection{Rabi Oscillations}

The two state system of Eq. (\ref{two-state-matrix-eq}) will Rabi oscillate between 
the $ \phi _{0} $ and $ \phi _{1} $
wavepackets in accordance with previous theory and experiments \cite{martin88,rabiosc}. 
The frequency at which each momentum state
oscillates between the two wavepackets is momentum dependent, and is given by the 
generalized Rabi frequency, defined as
\begin{equation}
\label{generalised-rabi-frequency}
\Omega ^{\prime }({\bf q})=\sqrt{V^{2}+\Delta _{0}({\bf q})^{2}.}
\end{equation}
%figure2
\begin{figure} 
\begin{center}
\epsfbox{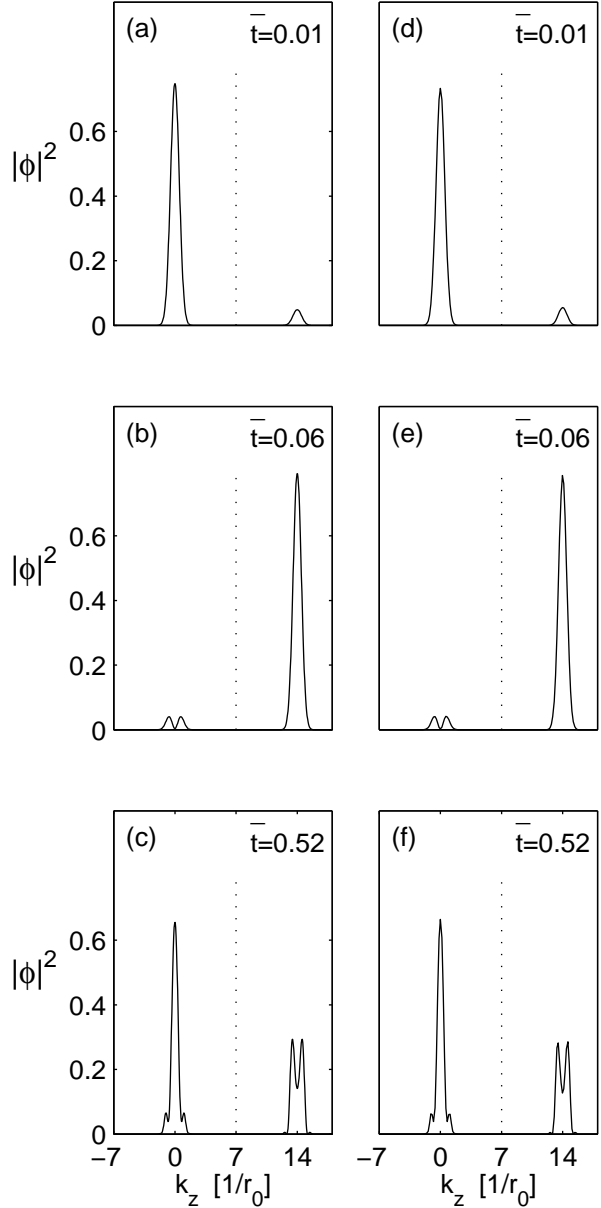}
\end{center}
\caption{\label{fig-rabi-osc-profiles}Evolution of the momentum
distribution for the two-state solution with $
\Delta k_{z}=14/r_{0}  $, $ \delta
=196\, \omega _{T}  $, and $ V=50\,
\omega _{T}  $.  The boundary for the
partitioned wavefunctions is indicated by the vertical dotted line. 
The wavepacket centred about $ k_{z}=0
 $ corresponds to $\phi_0$ and has an initial gaussian
profile.  The wavepacket to the right of the dotted line corresponds
to $\phi_1$ (but shifted by $ k_{z}=14/r_{0}
 $).  Figures (a)-(c): the analytic two-state solution. 
Figures (d)-(f): full numerical solution of Eq. 
(\ref{spatial-evolution-equation}) for $ w=0
 $.}
\end{figure}

 This oscillatory behaviour is clearly evident in Fig. 
\ref{fig-rabi-osc-profiles}(a)-(c), where we present the time
evolution of the momentum wavepackets from the two-state model, for a
case similar to Fig.  \ref{complex_density_fig}, but with $ w=0 $. 
As with all examples presented in sections
\ref{formalism}-\ref{twostate}, the condensate is prepared in an
eigenstate of a spherical harmonic trap and at time $ t=0 $ the trap
is switched off and the light grating is applied.  The initial state
is thus gaussian (since $ w=0 $), and we have also chosen the Bragg
detuning at the centre of the $ \phi _{0} $ wavepacket to be zero
(i.e. $ \Delta _{0}(0)=0 $ ).  In Fig. 
\ref{fig-rabi-osc-profiles}(b), where $ t=\pi /\Omega ^{\prime }(0)
$, the population has transferred almost entirely to the $ \phi _{1}
$ wavepacket.  The $ {\bf q} $ dependence of the Rabi cycling
becomes evident at later times as the difference in the Rabi periods
accumulates, an effect which can be seen in (Fig. 
\ref{fig-rabi-osc-profiles}(c)) where the cycling at the outer edges
of the momentum packets now significantly leads the cycling at $ {\bf
q}=0 $ giving rise to a pronounced central dip in $ \phi _{1} $. 
The full numerical solution of Eq. (\ref{spatial-evolution-equation})
for these parameters is presented in Fig. 
\ref{fig-rabi-osc-profiles}(d)-(f), and clearly confirms the
validity of the two-state model in this case.

\subsection{Dispersion curves and resonance-coupling}

The two-state model represented by Eq.  (\ref{two-state-matrix-eq})
was obtained by assuming that only one of the couplings in the matrix
of Eq.  (\ref{matrix_evolution_eq}) was important.  Here we present a
simple means of visualising the possible couplings, and determining
the validity and manner of breakdown of the two-state approximation.

%figure3
\begin{figure} 
\begin{center}
\epsfbox{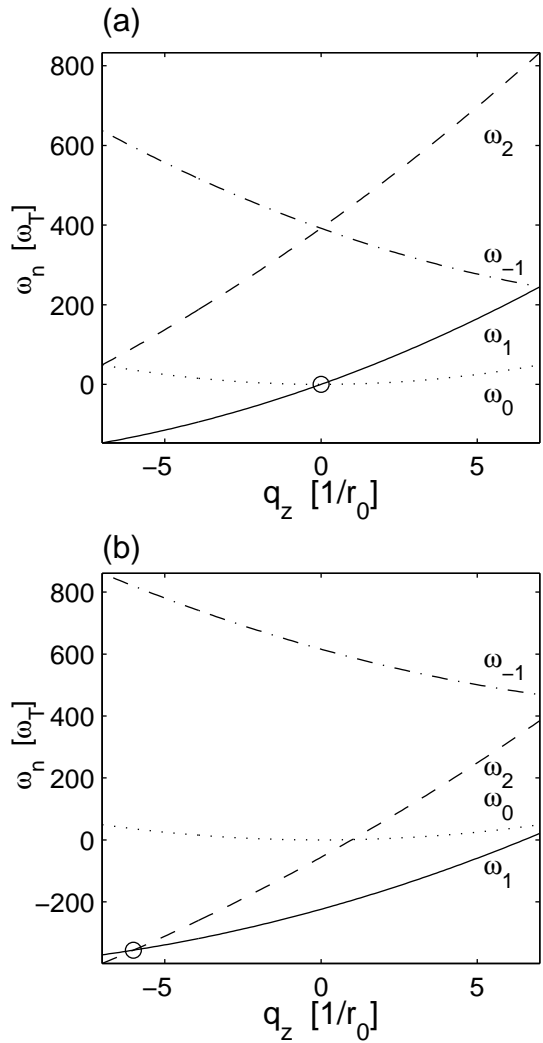}
\end{center}
\caption{\label{omega_curves_Fig} Free particle dispersion curves,
with first order resonance indicated by circle.  (a)
$ \Delta k_{z}=14/r_{0}  $ and
$ \delta =196\, \omega _{T}  $, (b)
$ \Delta k_{z}=14/r_{0}  $ and
$ \delta =420\, \omega _{T}  $ .  }
\end{figure}

We first construct a set of dispersion-type curves by plotting the
frequencies $ \omega _{n}({\bf q}) $ against $ {\bf q} $ for a
given choice of $ \Delta {\bf k} $ and $ \delta $, as illustrated
in Fig.  \ref{omega_curves_Fig}(a).  The important feature of this
graph is the vertical separation of the curves: the coupling between
$ \phi _{n} $ and $ \phi _{n+1} $ becomes appreciable only when $
\Delta _{n}({\bf q}) $ is small, i.e. when the difference between $
\omega _{n}({\bf q}) $ and $ \omega _{n+1}({\bf q}) $ is small.  We
need only consider $\Delta({\bf q})$ in the case of $ {\bf q} $ in
the $ z $ direction (the direction of $ \Delta {\bf k} $), because
the perpendicular components of $ {\bf q} $ simply offset all the $
\omega $ curves by the same amount.  We will therefore simply write
$q_z$ and $k_z$ for the momentum arguments in the remained of the
paper.  On the plots, the first order Bragg resonance condition ($
\Delta _{n}(q_{z})=0 $) appears as a crossing of the $ \omega _{n}
$ and $ \omega _{n+1} $ curves, and we have highlighted this point
(for $n=0$) with a small circle.  Appreciable coupling will only occur
in the vicinity of this point.  In Fig.  \ref{omega_curves_Fig}(a) we
have chosen the light grating parameters ($ \Delta {\bf k} $ and $
\delta $) so that the crossing is at $ q_{z}=0 $, and thus
wavepacket momentum components near $ q_{z}=0 $ will undergo Rabi
cycling.  In Fig.  \ref{omega_curves_Fig}(b), the grating parameter
choice puts the Bragg resonance at $ q_{z}=-6/r_{0} $, and the
response for a wavepacket centered at $ q_{z}=0 $ depends on the
coupling width, which we discuss in the following subsection.

Fig.  \ref{omega_curves_Fig}(b) also shows a crossing of the $ \omega
_{0} $ and $ \omega _{2} $ curves near $ q_{z}=0 $.  This is a
second order resonance: $ \phi _{0} $ can link to $ \phi _{2} $
via the intermediate state $ \phi _{1} $, and the overall process of
$ \phi _{0}\rightarrow \phi _{2} $ conserves energy and momentum. 
However the transition from $ \phi _{0}\rightarrow \phi _{1} $ (and
likewise $ \phi _{1}\rightarrow \phi _{2} $) is detuned by the
amount $ \Delta _{0}=36\omega _{T} $ at this point, and thus in
order that the transition $ \phi _{0}\rightarrow \phi _{2} $ may
proceed, $ V $ must be sufficiently large  for $ \phi _{1} $ 
to be seeded with population and
so mediate the resonant second order process.  This of course
generalises to higher order couplings.

\subsection{Coupling width and breakdown of two-state model}
\label{2statebreakdown}
As is well known from the Rabi model, the probability of transition
from state 1 to state 2 is given by the expression
\begin{equation}
\label{max_rabi_amp_eqn}
P_{12}=\frac{V^{2}}{V^{2}+\Delta _{0}(q_{z})^{2}}.
\end{equation}
As a function of $ q_{z} $, this is maximum at the first order Bragg
resonance $ \Delta _{0}(q_{z})=0 $ and falls to half its value at $
\Delta _{0}(q_{z})=V $, the power broadened width.  Momentum
components in $ \phi _{0} $ and $ \phi _{1} $ (or more generally
$ \phi _{n} $ and $ \phi _{n+1} $) will be significantly coupled
only if the separation of the $ \omega $ curves is less than $ V
$.  Converting this to the corresponding $ q_{z} $ width, we find
that the momentum width of the transition (i.e. the range of $ q_{z}
$ about the Bragg resonance point for which the probability of momentum
transfer exceeds $1/2$) is
\begin{equation}
\label{momentum-line-width-eq}
\sigma _{k}=\frac{2mV}{\hbar |\Delta k_{z}|}.
\end{equation}
The cycling behaviour of the whole wavepacket depends on the relative
size of the momentum width of the initial state, $ \delta k_{w} $,
compared to $ \sigma _{k} $ .  If $ \sigma _{k}>\delta k_{w} $ the
whole packet will cycle, as shown in Fig. 
\ref{fig-rabi-osc-profiles}. In practice this can be achieved using a
sufficiently large laser intensity (i.e. large $ V $ ) to broaden
the transition width $ \sigma _{k} $.  It is then possible to apply a $ \pi
$-pulse so that the initial momentum wavepacket ($ \phi _{0} $) is
transferred entirely to the adjacent location in momentum space ($
\phi _{1} $), and then the whole wavepacket receives a quantised
momentum change. A $\pi/2$ pulse on the other hand acts as a $50/50$ 
beam splitter.   For the case of $
\sigma _{k}\ll \delta k_{w} $, the Bragg process will couple out only
a fraction of the initial state, as we discuss further in Sec. 
\ref{applicationssec}.

%figure4
\begin{figure} 
\begin{center}
\epsfbox{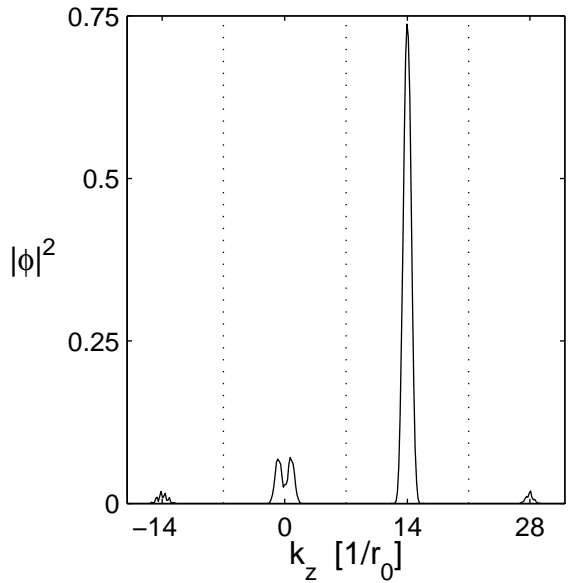}
\end{center}
\caption{\label{multiple_couplings_Fig} Numerical simulation of Bragg
scattering for large $ V  $, showing
probability distribution of $ z 
$-component of momentum at time $ t=0.29/\omega
_{T}  $.  The dotted lines indicate the boundaries
between the momentum partitions.  Light field parameters are
$ V=100\, \omega _{T}  $ ,
$ w=0  $, $ \Delta
k=14/r_{0}  $ and $ \delta =196\,
\omega _{T}  $.}
\end{figure}

From the dispersion curves in Fig.  \ref{omega_curves_Fig} it is
apparent that the validity of the two-state model requires that over the
momentum width $ \delta k_{w} $ of the initial wavepacket,
\emph{only} the $ \omega _{1} $ curve is within a distance $ V $
of $ \omega _{0} $ .  If on the other hand $ V $ is sufficiently
large that $ \omega _{-1} $ and $ \omega _{0} $, or $ \omega _{2}
$ and $ \omega _{1} $ are separated by less than (or of order $ V
$), then additional couplings will occur, and the two-state
description will fail.  We illustrate this in Fig. 
\ref{multiple_couplings_Fig}, where the parameters are the same as
in Fig.  \ref{fig-rabi-osc-profiles}, except that $ V $ has been
increased, to a value of approximately 25\% of the $ \omega _{1} $
to $ \omega _{2} $ separation.  By the time $ t=0.29/\omega _{T}
$, as shown in Fig.  \ref{multiple_couplings_Fig}, the additional
couplings have resulted in appreciable population transfer to the $
\phi _{-1} $ and $ \phi _{2} $ components.  Of course coupling to
additional $ \phi _{n} $ will occur if $ V $ is further increased. 
In view of this discussion we can formulate a condition that the
behaviour of a scattered wavepacket will be two-state in momentum
space: the minimum detuning to subsequent dispersion curves over the
momentum range of the wavepacket must be much greater than $V$, i.e.

\begin{equation}
\frac{\hbar }{m}\left[ |\Delta {\bf k}|^{2}-\frac{|\Delta {\bf k}|\,
\delta k_{w}}{2}\right] \gg V.
\end{equation}

It is worth emphasizing that for the case of a first order Bragg
resonance  the accuracy of the two
state model improves as $ \Delta {\bf k} $ increases in magnitude.

\section{Application to Nonlinear Case }
\label{appnonlinear}
In this section we compare our analytic results obtained in the linear
regime ($ w=0 $) to the case of nonzero values of  $ w $, and
characterise the regime where the linear analysis accurately describes
the nonlinear case.

\subsection{Numerical Result}

We first demonstrate that the main characteristics contained in the
two-state analysis still occur in the nonlinear case.

\begin{figure} 
\begin{center}
\epsfbox{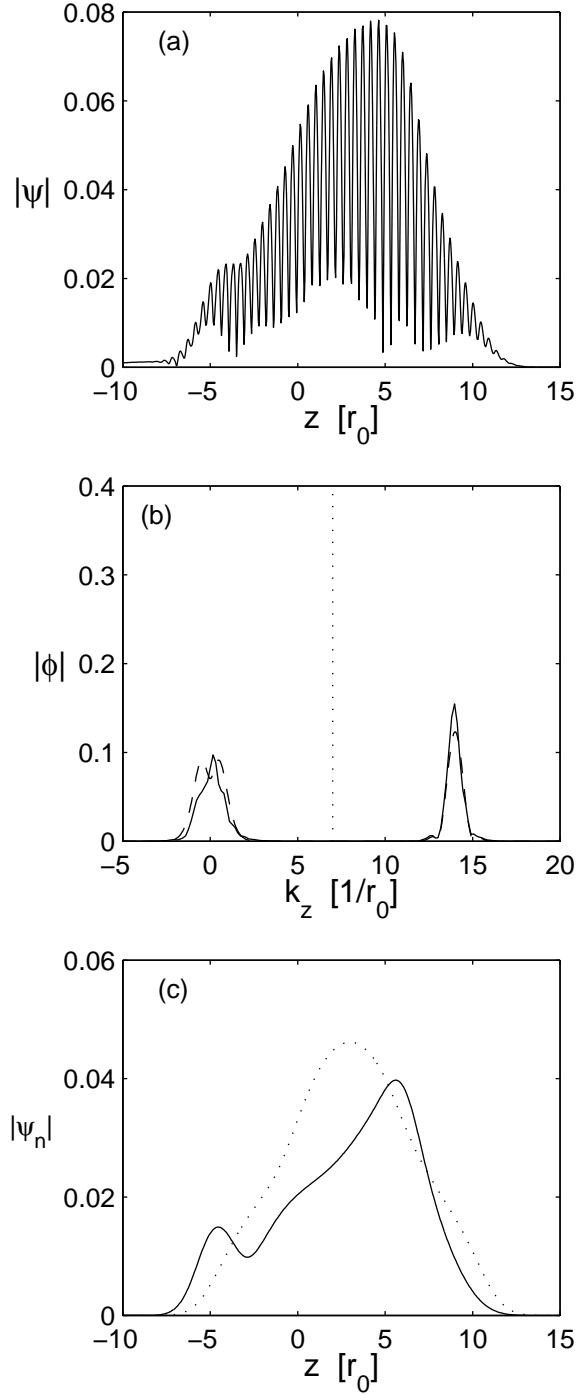}
\end{center}
\caption{\label{Spatial_decomposition_FIG}Spatial and momentum
profiles for the numerical solution presented in Fig. 
\ref{complex_density_fig} at $ t=0.21/\omega
_{T}  $.  (a) Spatial wavefunction amplitude along
the $ z  $-axis.  (b) Momentum
wavefunction amplitudes. Solid line corresponds to case in Fig. (a). Dotted line shows
the $ w=0  $ two-state case, 
for comparison.  (c) Amplitudes of $
\psi _{0}  $ (solid) and $ \psi
_{1}  $ (dotted) spatial wavefunctions corresponding
to $ \phi _{0}  $ and
$ \phi _{1}  $ of Fig. (b),
respectively.}
\end{figure}
We consider as an illustration the case presented in Fig. 
\ref{complex_density_fig}, where $ w=2500\, w_{0} $, and we plot in
Fig.  \ref{Spatial_decomposition_FIG}(a) the wavefunction amplitude
along the symmetry axis, corresponding to Fig.  1(b).  The
corresponding momentum wavefunction, obtained by numerical Fourier
transform, is shown in Fig.  \ref{Spatial_decomposition_FIG}(b), and
reveals two distinct packets.  On the same figure we have drawn for
comparison a dashed line representing the momentum wavefunction that
would arise from the $ w=0 $ two-state case.  The agreement is
close, and furthermore as time progresses the momentum wavepackets of
the $ w=2500\, w_{0} $ case oscillate at a frequency approximately
equal to the frequency $ V $ predicted by the two state
model.

The momentum distribution enables us to understand the fringe
structure that developed in Fig.  \ref{Spatial_decomposition_FIG}(a). 
We illustrate this by constructing separately the individual spatial
wavepackets $ \psi _{0} $ and $ \psi _{1} $ corresponding to the
momentum packets $ \phi _{0} $ and $ \phi _{1} $ respectively. 
The wavefunction $ \psi _{0} $ (shown as a solid line in Fig. 
\ref{Spatial_decomposition_FIG}(c)),
is of course essentially
stationary (apart from the effects of expansion), while the  
wavefunction $ \psi _{1} $ (shown as a dotted line) has a mean momentum of $ \hbar
k_{z}=\hbar \Delta k $, and moves to the right at a speed of $
\hbar \Delta k/m $.  This packet accordingly has a steep,
approximately linear, phase gradient across it, so that superposition
of $ \psi _{0} $ and $ \psi _{1} $ results in the observed
interference fringes.

\subsection{Free Expansion}

\begin{figure} 
\begin{center}
\epsfbox{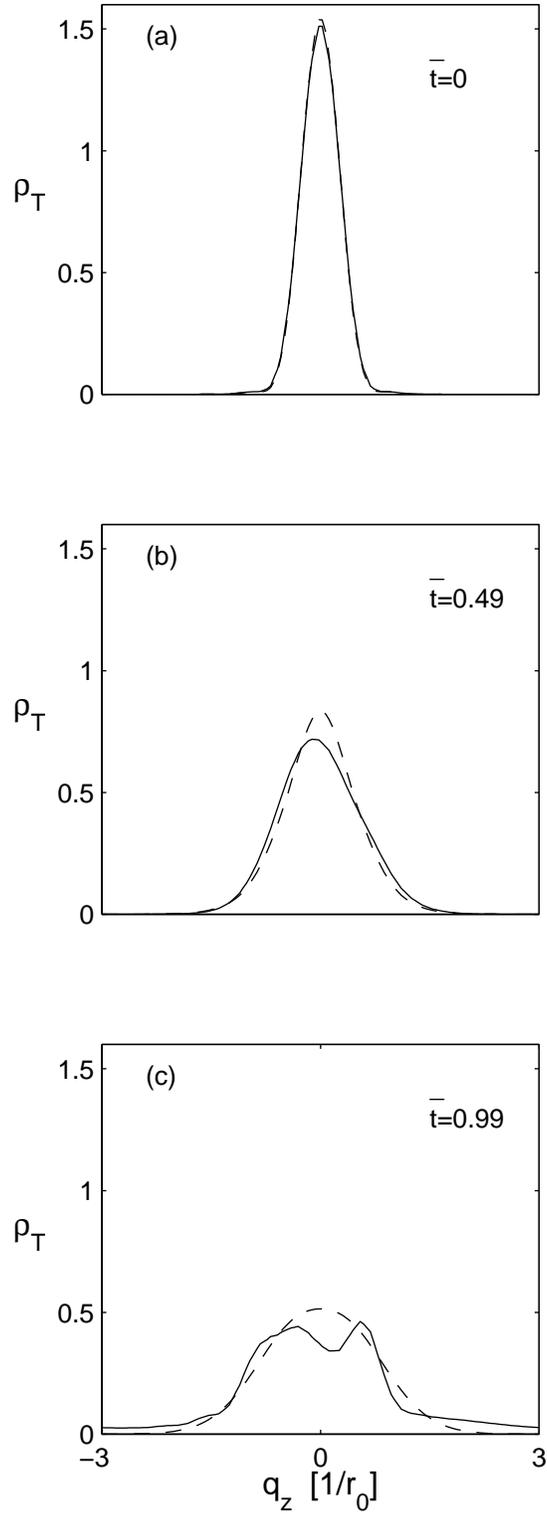}
\end{center}
\caption{\label{Ballistic_exp_FIG} Comparison between a purely
ballistically expanding condensate (dashed line) and a Bragg scattered,
untrapped condensate (solid line).  In each case $ w=2\,
500\, w_{0}  $.  The light field parameters for the
Bragg scattered state are $ V=5\omega_T  $,
$ \delta  =36\omega _{T}  $
and $ \Delta k=36/r_{0}  $.}
\end{figure}

One of the significant new effects in nonlinear Bragg scattering is that the
self repulsion in the free expansion of the condensate causes the
momentum wavepackets to expand.  In order to clearly demonstrate this
effect, it is convenient to isolate the contribution to momentum
changes that arise simply from the Bragg kicks.  We achieve this by
defining a momentum density
\begin{equation}
\rho _{T}({\bf q})=\sum _{i}|\phi _{i}({\bf q})|^{2},
\end{equation}
which gives the total occupation of all momentum states that could be
reached from an initial state of momentum $ \hbar {\bf q} $ by an
integral number of momentum kicks ($ n\hbar \Delta k $).  In the
linear case, where Bragg scattering is the only mechanism for changing
momentum states, $ \rho_T({\bf q}) $ is time independent.  In a
nonlinear condensate the effect of the ballistic expansion alone can
be seen in Fig.  \ref{Ballistic_exp_FIG}, where the dashed line
shows in successive frames, the momentum expansion of a $ w=2500\,
w_{0} $ cylindrically symmetric condensate freely evolving after release from a trap.  The
solid line shows the evolution of the same condensate in the presence
of Bragg scattering.  One sees that most of the momentum reshaping of
$ \rho_T ({\bf q}) $ is due to the repulsive expansion which occurs
on a time scale
\begin{equation}
    \label{time-const}
    \tau_E={\hbar \over \mu}{7\sqrt{3}\over 4},
\end{equation}
where $\mu$ is the chemical potential (see Appendix \ref{timeconstsec}).  For times $
t<\tau _{E} $ (see Table 1), we can neglect the effects of ballistic expansion, and
the effects of the condensate nonlinearity are confined to
modifications of the resonance conditions, which we consider in the
next section.

%figure6

\subsection{Nonlinear Dispersion Curves}
\label{nonlin_disp_curve}
The other significant modification arising from condensate nonlinearity
occurs in the Bragg resonance conditions, and the dispersion curve
concept provides a useful framework for analysing this effect.  In the
linear case, the diagonalisation of the Hamiltonian to determine the
energy eigenvalues and their momentum dependence is trivial, and leads
to the dispersion curves defined by Eq.  (\ref{omega-eq}), and
illustrated in Fig.  \ref{omega_curves_Fig}.  In the nonlinear case
diagonalisation is rather less straightforward: for small population
transfer from the ground state the appropriate dispersion curves are
given (for the case of Bragg scattering in a trap) by the Bogoliubov
dispersion relation \cite{bogoliubov}, however for large population
transfer the dispersion curves have not been investigated. 
Nevertheless as $ w $ increases from zero we can estimate that the
dispersion curves will be shifted by an amount of order $ \mu $, the
chemical potential.

\begin{figure} 
\begin{center}
\epsfbox{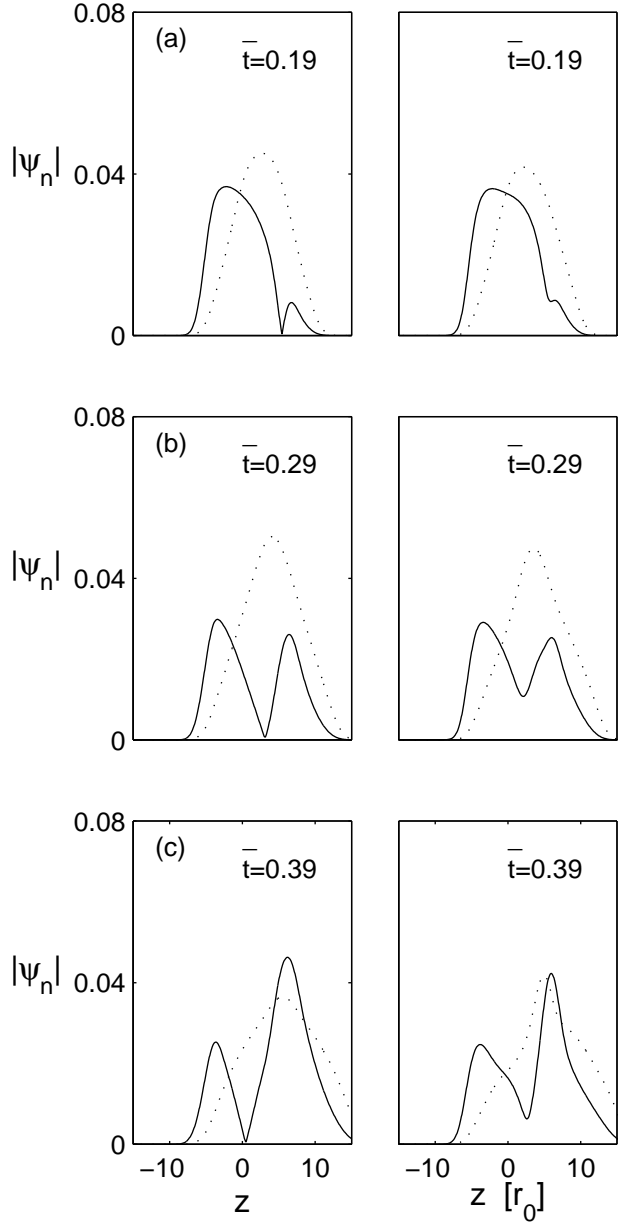}
\end{center}
\caption{\label{leap_frog_FIG} Comparison of Bragg scattered spatial
wavefunctions in linear and nonlinear condensates.  The solid curve
represents $ \psi _{0}  $ and the
dotted curve $ \psi _{1}  $.  
(a)-(c): $
w=0.  $ (d)-(f): $ w=2500\,
w_{0}  $.  Light field parameters are
$ V=10\omega _{T}  $,
$ \delta =196\omega _{T}  $,
$ \Delta k_{z}=14/r_{0}  $.}
\end{figure}

In our analysis of the linear case in the previous section, Bragg
scattering was shown to occur where the $ \omega _{n} $ and $
\omega _{n+1} $ curves are separated by less than the power broadened
width $ V $ of the laser interaction.  Thus if the laser parameters
$ \delta $ and $ \Delta k $ are chosen so that $ \omega _{0} $
and $ \omega _{1} $ cross in the $ w=0 $ case, we can expect our
linear analysis to still apply when $ w\ne 0 $, provided the shift
of the dispersion curves, $ \mu $, is less that the transition
width.  The regime where the results from section \ref{twostate} can
be applied in the nonlinear case is therefore
\begin{equation}
\label{nl_cond_EQ}
\mu\ll \hbar V,
\end{equation}
and we list in Table 1 the values of the $ \mu$ at the instant of
release from the trap for a range of condensates. We note that if $V$ becomes
sufficiently large, Bragg scattering will evolve into quantum channeling
\cite{channeling}.
 A critical test of
condition (\ref{nl_cond_EQ}) can be made by considering condensate
evolution in the spatial picture, and the comparison between the
linear and nonlinear cases is facilitated by separating the spatial
wavefunctions into their $ \psi _{0} $ and $ \psi _{1} $
constituents (corresponding respectively to $ \phi _{0} $ and $
\phi _{1} $).  We have plotted a temporal progression of these
wavefunctions in Fig.  \ref{leap_frog_FIG} for the $ w=0 $ case
(Figs.  \ref{leap_frog_FIG}(a)-(c)) and for $ w=2500\, w_{0} $ (Figs.
\ref{leap_frog_FIG}(d)-(f)).  For the latter case we have $
\mu=8.26\hbar \omega _{T} $ and $ V=10\omega _{T} $, so that
condition (\ref{nl_cond_EQ}) is satisfied.  It is evident that there
is very good agreement between the linear and nonlinear cases in all
the qualitative features, with the main difference perhaps being the
smoothing of the sharpest features in the nonlinear case.

%figure7

\begin{figure} 
\begin{center}
\epsfbox{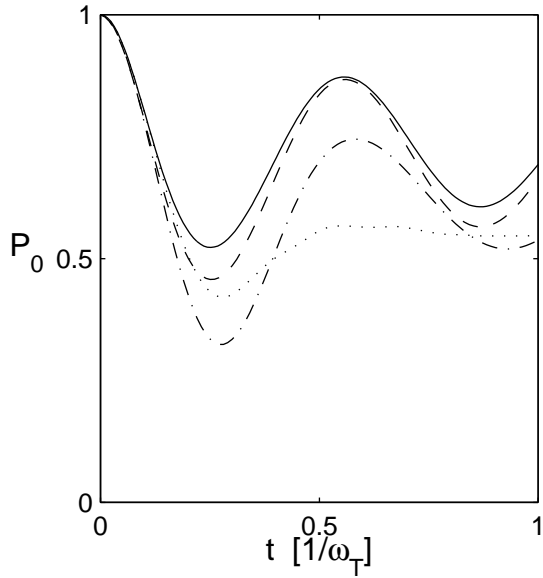}
\end{center}
\caption{\label{nl_pop_rabi_FIG} Temporal evolution of the population
of the $ 0^{\hbox {th}}  $ momentum
domain, for different condensate nonlinearities.  $
w=0  $ solid, $ w=250
 $ dashed, $ w=2\, 500  $
dash-dot, and $ w=25\, 000  $
dotted.  Light field parameters are as in Fig. 
(\ref{leap_frog_FIG}).}
\end{figure}

Finally we illustrate in Fig.  \ref{nl_pop_rabi_FIG} how the linear
analysis progressively fails as $ \mu $ approaches and then exceeds
$ \hbar V $.  We plot for the range of $ w $ values given in Table 1 the
time evolution of, $P_0$, the total population in the $ 0^{\hbox
{th}} $ momentum domain, which is defined according to
\begin{equation}
\label{popdef}
P_{n}=\int dk_x \int dk_y \int _{-{\Delta k}/{2}}^{{\Delta k}/{2}}
dq_{z}|\phi _{n}({\bf q})|^{2}\,.
\end{equation}

The linear case is shown as a solid line, and is in close agreement
with the $ w=250w_{0} $ case, and in reasonable agreement with the
$ w=2500w_{0} $ case.  However for the $ w=25000w_{0} $ case,
where $ \mu $ significantly exceeds $ \hbar V $, significant
disagreement occurs after the first half cycle.  We note that by using
$ P_{0} $ rather that $ \phi _{0}(q_{z}) $ as our basis of
comparison, we have minimised the discrepancy that would occur simply
from the repulsive expansion.

%figure8

%table1
\medskip
\begin{center}
\begin{tabular}[c]{|c|c|c|}
    \hline
$ w \quad [w_0]$& $ \mu \quad [\hbar\omega_T]$& $ \tau_E \quad
[1/\omega_T]$\\
\hline
$ 0 $& $ 0 $& $ \infty $\\
\hline
$ 250 $& $ 3.814 $& $ 0.8 $\\
\hline
$ 2\, 500 $& $ 8.258 $& $ 0.37 $\\
\hline
$ 25\, 000 $& $ 20.42 $& $ 0.15 $ \\
\hline
\end{tabular}
\end{center}
\medskip \textbf{Table: 1} Chemical potentials and expansion time constants 
for the eigenstates used in section \ref{nonlin_disp_curve}.

\section{Momentum Spectroscopy}
\label{applicationssec}

\begin{figure} 
\begin{center}
\epsfbox{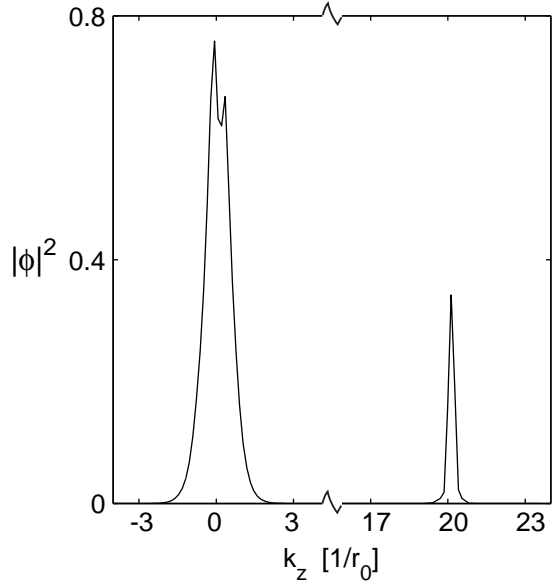}
\end{center}
\caption{\label{Momentum_Nibble_Fig} Momentum spectroscopy.  The
momentum wavefunction along the symmetry axis of the condensate is
shown at $ t=0.5/\omega _{T}.  $ The
out coupled state (centred at $k_z=20/r_0$) is much narrower than
the initial packet which is centred at $
k_{z}=0  $.  Parameters are $
w=2500\, w_{0}  $, $ V=5 \omega
_{T}  $, $ \Delta k=20/r_{0}
 $ and $ \delta =400\omega _{T}
 $.  }
\end{figure}

The NIST and MIT groups recently reported experiments
\cite{kozuma99}, \cite{stenger99} where Bragg scattering was used to
directly measure the momentum composition of condensates.  The
formalism set out in this paper provides an appropriate framework for
the description of those results.  Bragg scattering can be used to
selectively couple out a portion of the momentum states from a
wavepacket, provided the momentum width $ \sigma _{k} $ of the Bragg
transition is less than the momentum width $ \delta k_{w} $ of the
initial wavepacket.  The width $ \sigma _{k} $ is controlled
primarily by the laser intensity, and it is easy to see from Eq. 
(\ref{momentum-line-width-eq}) that for the purposes of momentum
spectroscopy, we require the laser intensity to be chosen such that

\begin{equation}
\label{V_codn_moment_spec_EQ}
V<(\hbar |\Delta {\bf k}|/4m)\: \delta k_{w}.
\end{equation}
 We illustrate such a case in Fig.  \ref{Momentum_Nibble_Fig} where a
 low intensity, high momentum coupling~ causes a narrow fraction of
 the initial momentum distribution~ to be out-coupled.

%figure9

\begin{figure} 
\begin{center}
\epsfbox{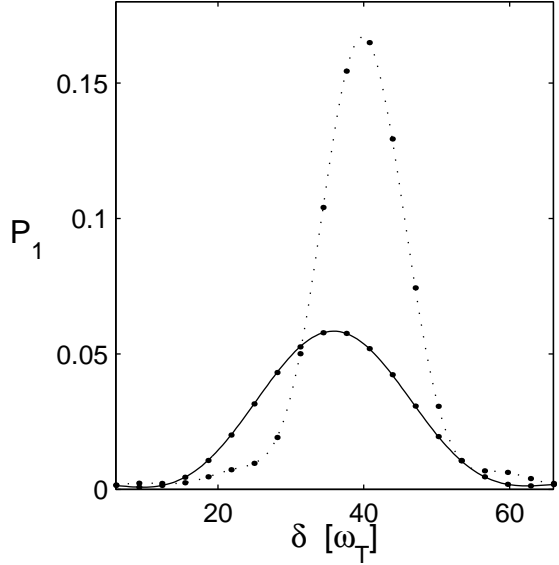}
\end{center}
\caption{\label{nonlinear_shift_Fig.}The total population $P_1$ (as
defined in Eq.  (\ref{popdef})) coupled into the $
\phi _{1}  $ wavepacket as a function of $\delta$ for
a $ w=0  $ condensate (solid line)
and a $ w=2500\, w_{0}  $ condensate
(dotted) as a function of $ \delta  
$.  Parameters of the field are $ \Delta
k=6/r_{0}  $, $ V=2\omega
_{T}  $ and it is applied for a time
$ T=0.5/\omega _{T}  $, while the
harmonic trap still remains on.}
\end{figure}

In the experiments  reported by the NIST group,
momentum selectivity was obtained by allowing the condensate to ballistically
expand before the application of the light grating, so that the self
repulsion broadens the momentum distribution to be larger than the
Bragg line width $ \sigma _{k} $.  The MIT group achieve precise
momentum selectivity by using very low intensity lasers, which allows
them to analyse the narrow momentum distribution of a trapped
condensate.  The MIT group have also been able to measure the shift of
the resonant point of the Bragg transition, which was discussed in section
\ref{appnonlinear}, and is calculated to be
\cite{stamper99,zambelli99}, in the regime of their experiments
\begin{equation}
\label{Braggshift_EQ}
\Delta \delta =\frac{4\mu }{7\hbar}.
\end{equation}
Our simulations confirm this expression.  Fig. 
\ref{nonlinear_shift_Fig.} presents the results for the out-coupled
population in a sequence of simulations where the detuning $ \delta
$ of the two Bragg fields is scanned through a range, while the other
field parameters are kept constant.  The solid line joins the
simulation data points for the linear case ($ w=0 $), while the
dotted line joins the data points for the case $ w=2500w_{0} $.  In
each case, the Bragg pulse is applied for a time $ T=0.5/\omega _{T}
$ while the condensate is still in the trap.  The two curves have
different shapes because the initial momentum wavepackets are
different for the two cases, but the offset of the peaks, which
represents the nonlinear shift of the Bragg resonance is measured to
be $ \sim 5\omega _{T} $, which is in very good agreement with
Eq.  (\ref{Braggshift_EQ}).

%figure10

\section{Conclusion}

In this paper, we have presented in detail a meanfield formalism for Bragg
scattering in Bose condensates. Starting from the two-component
Gross-Pitaevskii equation driven by a far detuned classical light field, we
obtained a one-component equation which provides a framework
for treating the scattering of trapped or untrapped condensates from an
arbitrary light field grating. We have concentrated on the Bragg regime, and
solved the one component equation in a number of situations using three
dimensional cylindrically symmetric numerical simulations. For the case of
Bragg scattering from a condensate released from a trap, the dominating
behaviour is the oscillation of the atomic momentum components between an
initial value ${\bf k_{\rm i}}$ and a shifted value ${\bf k_{\rm i}} + \Delta
{\bf k}$. We presented an alternative derivation of linear Bragg scattering
based on a partitioned representation of the momentum wavefunction and
incorporating spatial nonuniformity, and used this to analyse the full
nonlinear system. We showed that this linear model gives an accurate
description of the nonlinear system within a well defined validity regime.
Dispersion curves in the partitioned momentum representation facilitated
visualisation of the possible couplings that may occur, including first and
higher order Bragg resonances. We also derived an analytic expressions for the
the momentum transition linewidth, and used this to define the regime of
momentum spectroscopy. Finally, we used numerical simulations to investigate
Bragg scattering from a condensate in an trap, and to calculate the nonlinear
shift of the Bragg resonance frequency.

\section*{Acknowledgments}
The authors thank K-P. Marzlin, C. W. Gardiner and K. Burnett for
helpful discussions, and M. J. Davis for his careful reading of the 
manuscript.  This work was supported by Marsden Grants PVT
603 and PVT 902.

\begin{appendix}
%APPENDIX A
\section{Adiabatic Elimination}
\label{adiaelimsec}
Eqs.  (\ref{full_evolve_eq}) and (\ref{full_evolve_EQ2}) detail the EM
field coupling, in the dipole approximation, between the meanfields in
the internal states $ |g\rangle $ and $ |e\rangle $ respectively. 
By defining a new excited meanfield amplitude
\begin{equation}
\tilde{\psi }_{e}({\bf r},t)=e^{i\omega _{1}t}\, \psi _{e}({\bf r},t),
\end{equation}
 the evolutions equations are transformed to
\begin{eqnarray}
i\hbar \frac{\partial \psi _{g}}{\partial t} & = & -\frac{\hbar
^{2}}{2m}\nabla ^{2}\psi _{g}+V_{Tg}({\bf r},t)\psi_g-\frac{1}{2}\hbar\Omega _{0}(t)\Lambda
({\bf r},t)e^{-i\omega _{1}t}\tilde{\psi }_{e}\label{rw_evolve_eqnA}
\\
 & + & (w_{gg}|\psi _{g}|^{2}+w_{eg}|\tilde{\psi }_{e}|^{2})\psi
 _{g},\nonumber \\
 & & \nonumber \\
i\hbar \frac{\partial \tilde{\psi }_{e}}{\partial t} & = &
-\frac{\hbar ^{2}}{2m}\nabla ^{2}\tilde{\psi }_{e}+V_{Te}({\bf r},t)\tilde{\psi}_e-\frac{1}{2}\hbar
\Omega ^{*}_{0}(t)\Lambda ({\bf r},t)e^{i\omega _{1}t}\psi
_{g}\label{rw_evolve_eqnB} \\
 & & -\hbar \Delta \psi _{e}+(w_{eg}|\psi _{g}|^{2}+w_{ee}|\tilde{\psi
 }_{e}|^{2})\tilde{\psi} _{e},\nonumber
\end{eqnarray}
 where we have introduced the detuning $ \Delta =\omega _{1}-\omega
 _{eg} $.

For an initial configuration with the ground state fully occupied, we
make the approximations that in the region of significant condensate density  
the magnitude of $ \Delta \tilde{\psi
}_{e} $ greatly exceeds those of $ w_{ee}|\tilde{\psi
}_{e}|^{2}\tilde{\psi }_{e} $, $ {\hbar ^{2}\nabla ^{2}}\tilde{\psi
}_{e}/{2m} $, $ w_{eg}|\psi _{g}|^{2}\tilde{\psi }_{e} $ and $V_{Te}({\bf r},t)\tilde{\psi}_e$ 
(which can always be arranged by choosing $\Delta$ sufficiently large), so
that these latter three terms can be dropped from Eq. 
(\ref{rw_evolve_eqnB}) allowing the following approximate formal
solution

\begin{equation}
\label{formal_adia_elim_int_eq}
\tilde{\psi }_{e}({\bf r},t)\approx +\frac{i}{2}\int _{0}^{t}
\left \{e^{i\Delta (t-s)}\Lambda ^{*}({\bf r},s)e^{i\omega
_{1}s} \right \}\Omega _{0}^{*}(s)\psi _{g}({\bf r},s)\, ds.
\end{equation}

The adiabatic elimination of the excited state proceeds by noting that
while $ |\Omega _{0}|\ll |\Delta | $ and $|\Omega_0^\prime(t)|\ll|\Delta\,\Omega_0(t)|$,
 $\Omega_0^*(s)\psi _{g}({\bf r},s)$ is
much more slowly varying in time than the terms in the braces in the
integrand of Eq.  (\ref{formal_adia_elim_int_eq}), which vary at least
as fast as $ e^{i\Delta s} $. 
 This means the main contribution to
the integral arises near the end point, and so $ \Omega_0(s)\psi _{g}({\bf r},s)$
 may be taken outside the integral as $\Omega_0(t)\psi_{g}({\bf r},t)$, i.e.
\begin{equation} 
\label{adia_elim_eq}
\tilde{\psi }_{e}({\bf r},t)\approx \frac{i\Omega_0(t)}{2}\,\psi_{g}({\bf r},t)
e^{i\Delta t} \int_{0}^{t}e^{-i\Delta s} \left \{ e^{i{\bf k}_1\cdot{\bf r}}+
e^{i({\bf k}_2\cdot{\bf r}-(\omega_2-\omega_1)s)} 
+e^{-i({\bf k}_1\cdot{\bf r}-2\omega_1s)}
e^{-i({\bf k}_2\cdot{\bf r}-(\omega_2+\omega_1)s)} \right \}\,ds.
\end{equation}
 The
rotating wave approximation (RWA) is now made, in which we neglect the
last two rapidly varying terms in Eq. (\ref{adia_elim_eq}).  The resulting
approximate solution to Eq.  (\ref{formal_adia_elim_int_eq}) is

\begin{equation}
 \tilde{\psi }_{e}({\bf r},t)\approx -\frac{\Omega
 _{0}^{*}(t)}{2\Delta }(e^{i{\bf k}_{1}\cdot {\bf r}}+e^{i({\bf k}_{2}\cdot
 {\bf r}-(\omega_{2}-\omega_1)t)})\, \psi
 _{g}({\bf r},t), \end{equation} which can be substituted into Eq. 
 (\ref{rw_evolve_eqnA}) to give

\begin{eqnarray}
i\hbar \frac{\partial \psi _{g}}{\partial t} & = & -\frac{\hbar
^{2}\nabla ^{2}}{2m}\psi _{g}+\frac{\hbar |\Omega
_{0}(t)|^{2}}{4\Delta }(e^{i({\bf k}_{1}\cdot {\bf r}-\omega
_{1}t)}+e^{i({\bf k}_{2}\cdot {\bf r}-\omega _{2}t)})\Lambda ({\bf
r},t)\psi_g \label{Adia_elime_last_EQ} \\
 & & +w_{gg}|\psi _{g}|^{2}\psi _{g}.\nonumber
\end{eqnarray}
 The RWA is again be applied to Eq.  (\ref{Adia_elime_last_EQ}),
 rejecting terms oscillating at optical frequencies. This leads
 to an optical potential of the form
 \begin{equation}
 \label{opt_pot_with_DC}
  {\hbar|\Omega_0(t)|^2\over2\Delta}[1+\cos(\Delta{\bf k}\cdot{\bf r}-\delta t)].
  \end{equation}
  Finally the phase choice $ {\psi}=\exp(-i\int_0^tV(s)ds)\,\psi_g,$ 
  removes the D.C term from this potential and yields Eq. (\ref{spatial-evolution-equation}).

%APPENDIX B
\section{Momentum Expansion Time Constant}
\label{timeconstsec}
Here we derive the approximate expression (Eq.  (\ref{time-const}))
for the time constant of momentum space expansion of a BEC. To do this
we consider the transfer of self energy to kinetic energy for a freely
expanding spherically symmetric condensate in the Thomas-Fermi limit. 
It is convenient to do this calculation in dimensionless units (see
section \ref{dimensionless_unit_sec}).

Writing the full momentum wavefunction as

\begin{equation}
    \phi(k,t) = \varphi(k,t)e^{iS(k,t)},
\end{equation}
where $\varphi(k,t)$ and $S(k,t)$ are real, we then approximate the
amplitude $\varphi$ of the momentum space wavefunction as a gaussian,

\begin{equation}
    \label{momenum-amp-defn}
    \varphi(k,t)=\left( \frac{3}{2\pi \sigma ^{2}}\right)
    ^{\frac{3}{4}}e^{-\frac{3k^{2}}{4\sigma ^{2}}}.
\end{equation}
The time dependence of $\varphi$ is contained in the function
$\sigma(t)$, the standard deviation of $\varphi^2$ (i.e $|\phi|^2$). 
The kinetic energy of $\phi$ depends only on $\varphi$, and can be
evaluated explicitly to give
\begin{equation}
\label{ke_var}
\langle KE\rangle =\int _{0}^{\infty }4\pi k^{2}\, k^{2}|\phi
(k,t)|^{2}dk=3\sigma(t) ^{2}.
\end{equation}
We shall use Eq.  (\ref{ke_var}) to attribute a momentum width to a
system whose kinetic energy is known.

During expansion, the total energy, consisting of the self energy and
the kinetic energy is conserved, so that the gain in kinetic energy
will be equal to the decrease in self energy.  We now proceed to
calculate the time dependent behaviour of the self energy.

At $ t=0 $ the harmonic trap is turned off and the initial condensate density
profile can be closely approximated by the Thomas-Fermi result
\begin{equation}
    \label{T-F-profile}
n_{TF}(r)=\frac{\mu -\frac{r^{2}}{4}}{w},
\end{equation}
 where $ 0<r<2\sqrt{\mu } $, and

\begin{equation}
\label{TF_mu_EQ}
\mu =\left( \frac{15w}{64\pi }\right) ^{\frac{2}{5}}.
\end{equation}

Castin and Dum \cite{castin96} have shown that in the Thomas-Fermi
approximation the spatial profile of a spherically symmetric released
BEC behaves as

\begin{equation}
n(r,t)=b^{-3}(t)\, n_{TF}(r/b(t)),
\end{equation}
 where
\begin{equation}
b(t)=\sqrt{1+t^{2}}.
\end{equation}

We use this expression to find the time dependent self energy,

\begin{eqnarray}
\langle SE\rangle (t) & = & \int ^{2b(t)\sqrt{\mu }}_{0}4\pi r^{2}\,
\left[ {w\over 2}\, n(r,t)^{2}\right] dr,\\
 & = & \frac{1}{14}\left( \frac{15w}{2\pi }\right)
 ^{\frac{2}{5}}(1+t^{2})^{-\frac{3}{2}},\label{SE_time_EQ}
\end{eqnarray}

and from this can from this we can write an expression for the KE as a function of
time,
\begin{equation}
    \label{change-in-KE}
    \langle KE\rangle(t) = \langle KE\rangle_i+\langle
    SE\rangle(0)-\langle SE\rangle (t).
\end{equation}

An approximation for the initial value of the kinetic energy $\langle
KE \rangle_i$ can be found by calculating the spatial wavefunction
corresponding to $\varphi$ and matching its variance to that of the
initial Thomas-Fermi profile of Eq.  (\ref{T-F-profile}).  This gives
us an estimate of $\sigma(0)$, which from Eq.  (\ref{ke_var}) yields
the initial kinetic energy
\begin{equation}
    \label{init-KE}
    \langle KE \rangle_i = \left({7\over 2}\right)\left({3\over
    2}\right)^{3\over 5}\left({\pi\over 5w}\right)^{2\over 5}.
\end{equation}
Now using Eq.  (\ref{change-in-KE}) and the expression $\langle
KE\rangle =3 \sigma^2(t)$ we obtain the time dependent momentum
variance
\begin{equation}
    \label{time-var}
    \sigma^2(t) = \left({7\over 6}\right)\left({3\over
    2}\right)^{3\over 5}\left({\pi\over 5w}\right)^{2\over 5} +
    \frac{1}{42}\left( \frac{15w}{2\pi }\right)
    ^{\frac{2}{5}}\left(1-(1+t^{2})^{-\frac{3}{2}}\right).
\end{equation}
Defining the time constant, $\tau_E$, to be the time it takes for the
width $\sigma$ to double, we obtain
\begin{equation}
\label{tau_full_EQ}
\tau_E =\frac{\sqrt{1-(1-\alpha )^{\frac{2}{3}}}}{(1-\alpha
)^{\frac{1}{3}}},
\end{equation}
 where
\begin{equation}
\alpha =147\left( \frac{3}{2}\right) ^{\frac{1}{5}}\left( \frac{\pi
}{5w}\right) ^{\frac{4}{5}}.
\end{equation}
This derivation is only valid when the Thomas-Fermi approximation is
valid, in which case $\alpha$ is a small parameter, and Eq. 
(\ref{tau_full_EQ}) can be expanded in inverse powers of $\mu$ (i.e
$\alpha^{1\over 2}$).  To first order in $\mu^{-1}$ this gives
\begin{equation}
\tau_E \approx \sqrt{\frac{2\alpha }{3}}=\frac{1}{\mu
}{\left(\frac{7\sqrt{3}}{4}\right).}
\end{equation}

%%%%%
 
\end{appendix}

\end{document}